\documentclass{article} 
\usepackage{main,times}

\usepackage{amsmath,amsfonts,bm}

\def\eqref#1{equation~\ref{#1}}

\def\1{\bm{1}}

\DeclareMathAlphabet{\mathsfit}{\encodingdefault}{\sfdefault}{m}{sl}
\SetMathAlphabet{\mathsfit}{bold}{\encodingdefault}{\sfdefault}{bx}{n}

\usepackage{graphicx}
\usepackage{subcaption}
\usepackage{siunitx}
\usepackage{bbm}
\usepackage{hyperref}
\usepackage{url}
\usepackage[capitalize,noabbrev]{cleveref}

\title{DiffObs: Generative Diffusion for Global\\Forecasting of Satellite Observations}

\author{Jason Stock\thanks{ Work done during an internship at NVIDIA.} \\
NVIDIA Corporation and Colorado State University\\
\texttt{stock@colostate.edu}
\AND
Jaideep Pathak, Yair Cohen, Mike Pritchard, Piyush Garg, Dale Durran,\\\textbf{Morteza Mardani \& Noah Brenowitz}\\
NVIDIA Corporation\\
\texttt{\{jpathak,yacohen,mpritchard,piyushg,ddurran,}\\\texttt{ mmardani,nbrenowitz\}@nvidia.com}
}

\iclrfinalcopy 
\begin{document}

\maketitle



\begin{abstract}
This work presents an autoregressive generative diffusion model (DiffObs) to predict the global evolution of daily precipitation, trained on a satellite observational product, and assessed with domain-specific diagnostics. The model is trained to probabilistically forecast day-ahead precipitation. Nonetheless, it is stable for multi-month rollouts, which reveal a qualitatively realistic superposition of convectively coupled wave modes in the tropics. Cross-spectral analysis confirms successful generation of low frequency variations associated with the Madden--Julian oscillation, which regulates most subseasonal to seasonal predictability in the observed atmosphere, and convectively coupled moist Kelvin waves with approximately correct dispersion relationships. Despite secondary issues and biases, the results affirm the potential for a next generation of global diffusion models trained on increasingly sparse, and increasingly direct and differentiated observations of the world, for practical applications in subseasonal and climate prediction. 
\end{abstract}

\section{Introduction}
As machine learning-driven global forecasting systems exit their infancy and move beyond weather \citep{pathak2022fourcastnet,bi2022pangu,chen2023fengwu,lam2022graphcast} toward climate \citep{watt2023ace,weyn2019can} timescales, whether they can be made to generate realistic convectively coupled tropical disturbances across daily to multi-week simulations becomes an important question. Such atmospheric variability has been a longstanding challenge to capture realistically in physics-based models \citep{randall2013beyond} and is still incompletely understood \citep{Zhangetal20}, yet regulates the subseasonal predictability of the Earth System, including important tropical to extratropical teleconnections \citep{lau2005predictability}. 

These dynamics become especially interesting to examine in emerging autoregressive diffusion models \citep{price2023gencast,mardani2024residual,li2023seeds,nath2023forecasting} that learn conditional probabilities and are thus well suited to the stochastic character of tropical convective dynamics. Moreover, such methods suggest that diffusion models \citep{karras2022elucidating,song2020score} do not require complete information about the atmospheric state, and thus may have the capacity to produce realistic variability even when trained on limited, direct (e.g., univariate) observations. Some recent work \citep{gao2023prediff, leinonen2023latent} has explored diffusion models with univariate weather data, but only on short-term scales and in small spatial domains; computational advances in GPU computing allow more ambition today.

In this context, we introduce a computationally ambitious, high-resolution ($0.4^\circ$) global autoregressive diffusion model trained solely on a satellite derived precipitation product. Precipitation is observed globally via microwave sensing satellites (e.g., TRMM, \cite{kummerow1998tropical} and GPM, \cite{hou2014global}), and prior low-order models have proven skillful at long-range forecasts \citep{Chen2014-jh}. The coupling of both data and our modeling approach is pivotal to the feasibility of our current work. Using our trained model, we perform an in depth analysis of generated tropical variability on 1- to 60-day timescale using domain-informed diagnostics, discovering long-term stability with realistic variability of multiple wave structures. 

\section{Methodology} \label{sec:method}
We introduce an autoregressive diffusion model, extending the EDM architecture \citep{karras2022elucidating} with the objective to estimate $p\left(\mathbf{x}_{t} \vert \mathbf{x}_{t-1}\right)$ without incorporating additional priors. Achieving this goal assumes a paired spatiotemporal relationship within the underlying distribution to effectively capture the dynamics of the system based solely on the immediate past state. In doing so, our model can rollout predictions by utilizing the estimated next step as the subsequent initial condition. 

The design specifics of our model are inspired by the work of prior global diffusion-based weather forecasting models. Specifically, we build upon the work of \cite{mardani2024residual}, which employs a similar architecture for \si{\kilo\meter}-scale downscaling. However, we avoid the use of an intermediate regression model and do not scale down the conditional inputs. \cite{price2023gencast} present a purely autoregressive diffusion model, but train on a comprehensive state vector given from reanalysis data, where instead we directly estimate a single observational state. Additional details in \cref{app:diffusion}.

\subsection{Training Details}
Our experiments use the default hyperparameters outlined in \cite{karras2022elucidating}, extending the DDPM++ UNet architecture \citep{song2020score}, with the only deviations being the exclusion of self-attention and a reduction in model channels, specifically from $128\rightarrow64$. Despite the limitations imposed on the model's receptive field and its ability to capture global synoptic information without self-attention, we find the change is needed to achieve reasonable performance and training stability. 

We train our $13.6$M parameter model on a cluster with $256\times$ \SI{80}{\giga\byte} H100 NVIDIA GPUs (32 nodes) using a global batch size of $1{,}024$ for $12.5$M total steps. End-to-end training takes \SI{4}{\hour} wall-clock time. During generation, we sample with $64$ denoising steps using the default noise levels. A single output image is fully generated in \SI{8.5}{\second} (unoptimized) using $4.6$ \si{\giga\byte} of memory on a single GPU.

\begin{figure*}[!t]
\centering
\includegraphics[width=1\textwidth]{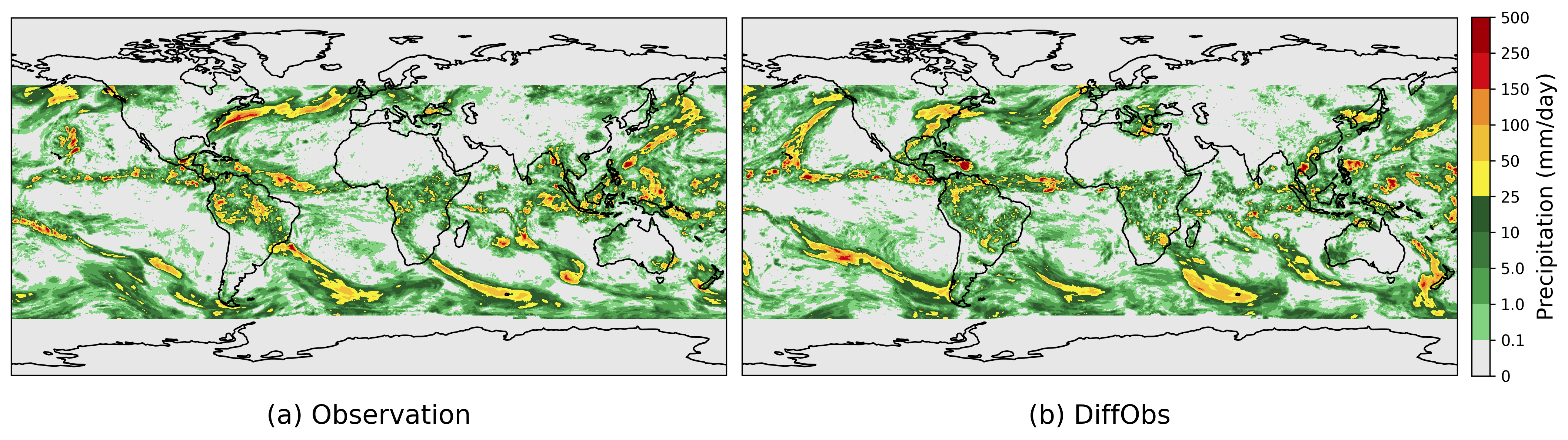} 
\caption{Example 3-day rollout from Oct 27, 2020 as initial condition.}
\label{fig:samples}
\end{figure*}

\section{Experiments} \label{sec:experiments}

We begin by introducing the dataset for this study in \cref{sec:dataset}, detailing the preprocessing and partitioning specifics, and in \cref{sec:results} we showcase and discuss our primary results.

\subsection{Dataset Details} \label{sec:dataset}
This study uses the final precipitation, half hourly Integrated Multi-satellitE Retrievals for Global Perception Measurements (IMERG) L3 Version 06B data \citep{huffman2019gpm,huffman2015nasa}. Global estimates are derived through intercalibration, morphing, and interpolating various satellite microwave precipitation and infrared retrievals, precipitation gauge analyses, and surface observations (e.g., temperature, pressure, and humidity). 

We collect data from June 1, 2000 to Sept 30, 2021 and aggregate all half hour samples for each day into an estimate of total daily precipitation (in \si{\milli\meter/\day}). Thereafter, we spatially coarsen the grid from $0.1^\circ \rightarrow 0.4^\circ$ with cropping in the meridional direction between $56.2^\circ$N and $61.8^\circ$S (296 latitudes and 900 longitudes) to avoid masking missing values at the poles. Data are partitioned to the years of 2000--2016 ($6{,}041$) for training and 2017--2022 ($1{,}729$) for testing, with the total samples in parentheses. Individual sample pairs, $\left(\mathbf{x}_t, \mathbf{x}_{t-1}\right)$, are on a one-day interval with $\mathbf{x}_{t-1}$ being the condition.

Even with daily-accumulated estimates, the distribution of data is heavily right-skewed and primarily comprised of zero-valued cells with few high, yet critical precipitation values (e.g., in locations with severe weather). We therefore transform the data to be relatively Gaussian, and using statistics from the training data, normalize it between $[-1,1]$ to align with the assumptions of diffusion models. This is done as $g(x) = 2\cdot \ln \left(1 + x/\epsilon\right) / x_{max} - 1$, where $\epsilon = 10^{-4}$ and $x_{max}=17.35$ as found in the transformed data. Computing $g^{-1}(x)$ on model output returns the data to its original units.

\begin{figure}[t!]
    \centering
    \begin{subfigure}{0.49\textwidth}
        \includegraphics[width=\textwidth]{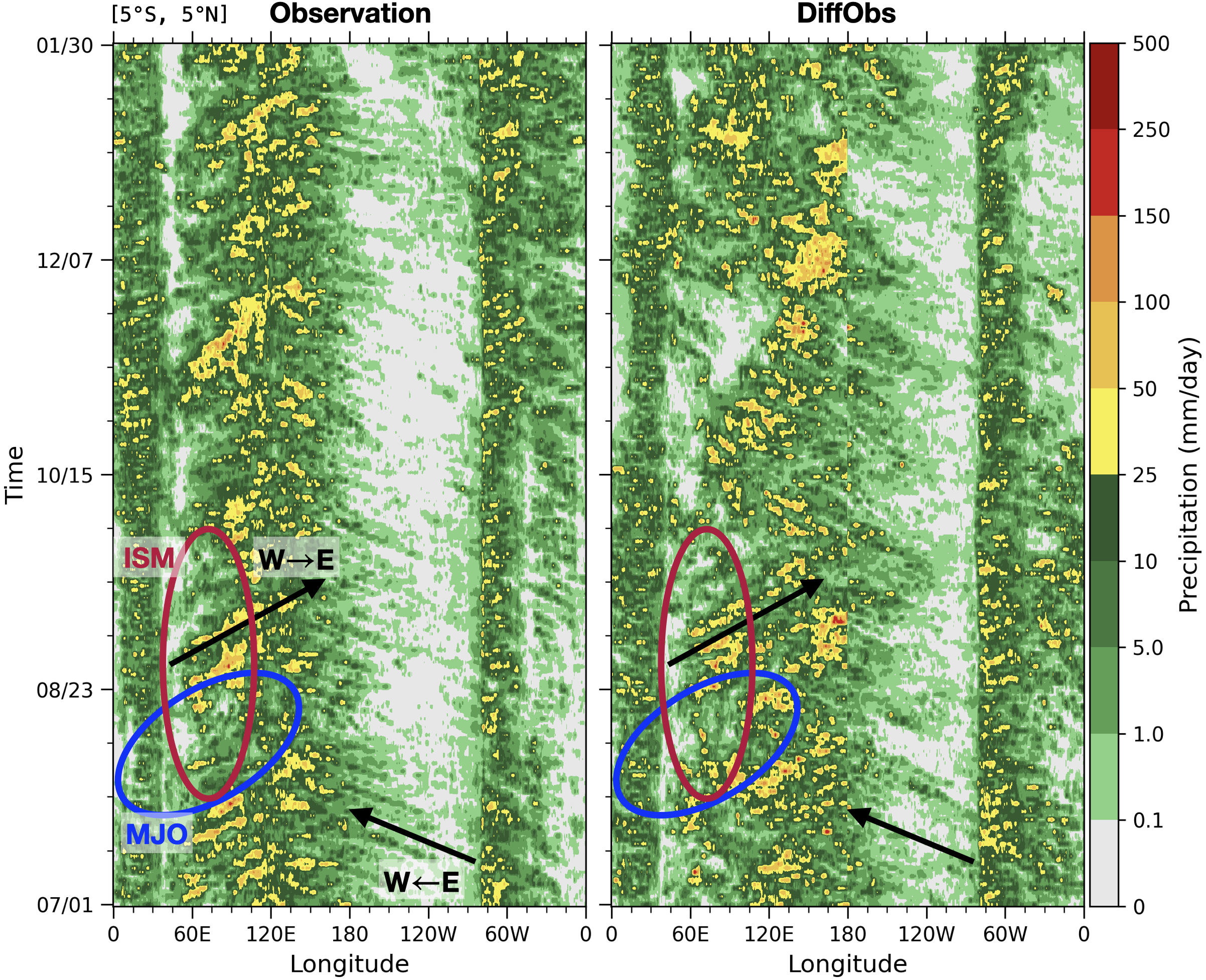}
        \caption{Summer to Winter}
        \label{fig:results-hov-a}
    \end{subfigure}
    ~ 
    \begin{subfigure}{0.49\textwidth}
        \includegraphics[width=\textwidth]{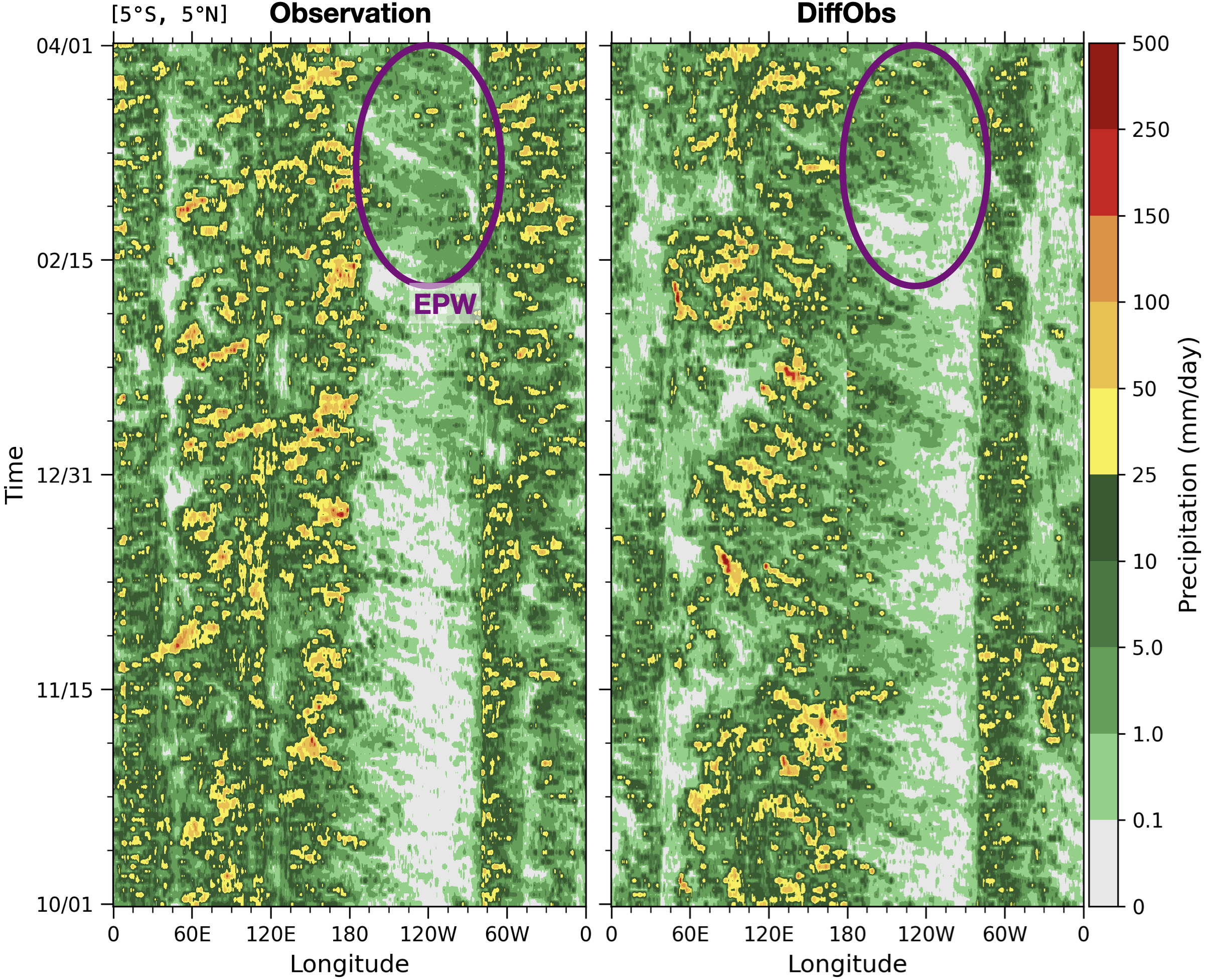}
        \caption{Boreal Winter}
        \label{fig:results-hov-b}
    \end{subfigure}
    \caption{Hovm\"{o}ller diagrams of observations (left) and DiffObs output (right, generated autoregressively) between $5^\circ$N and $5^\circ$S for case studies initially conditioned on (a) July 1, 2020 and (b) Oct 1, 2019. Individual colors correspond to the wave propagation directions (W$\leftrightarrow$E), Indian Summer Monsoon (\textcolor{purple}{ISM}), Madden‐-Julian oscillation (\textcolor{blue}{MJO}), and East Pacific Wavetrain (\textcolor{violet}{EPW}).}
    \label{fig:results-hov}
\end{figure}

\subsection{Main Findings} \label{sec:results}
Using our trained model, we can generate forecasts for arbitrary lead times and leverage its inherent probabilistic nature to create an ensemble of forecasts from any initial condition. A representative member of an ensemble is shown in \cref{fig:samples}, featuring a 3-day rollout initialized with a sample from Oct 27, 2020, to illustrate example output. While exact features should not be expected to match perfectly due to atmospheric chaos, it is notable that the forecast maintains qualitative sharpness, addressing concerns observed in deterministic convolutional networks \citep{ravuri2021skilful,ayzel2020rainnet}, and accurately captures the structure of atmospheric conditions, including many high-valued precipitation events. Next, we shift our focus to evaluate long, multi-month rollouts to study multi-scale generated atmospheric variability near the equator.

\begin{figure}[t!]
    \centering
    \begin{subfigure}{0.49\textwidth}
        \includegraphics[width=\textwidth]{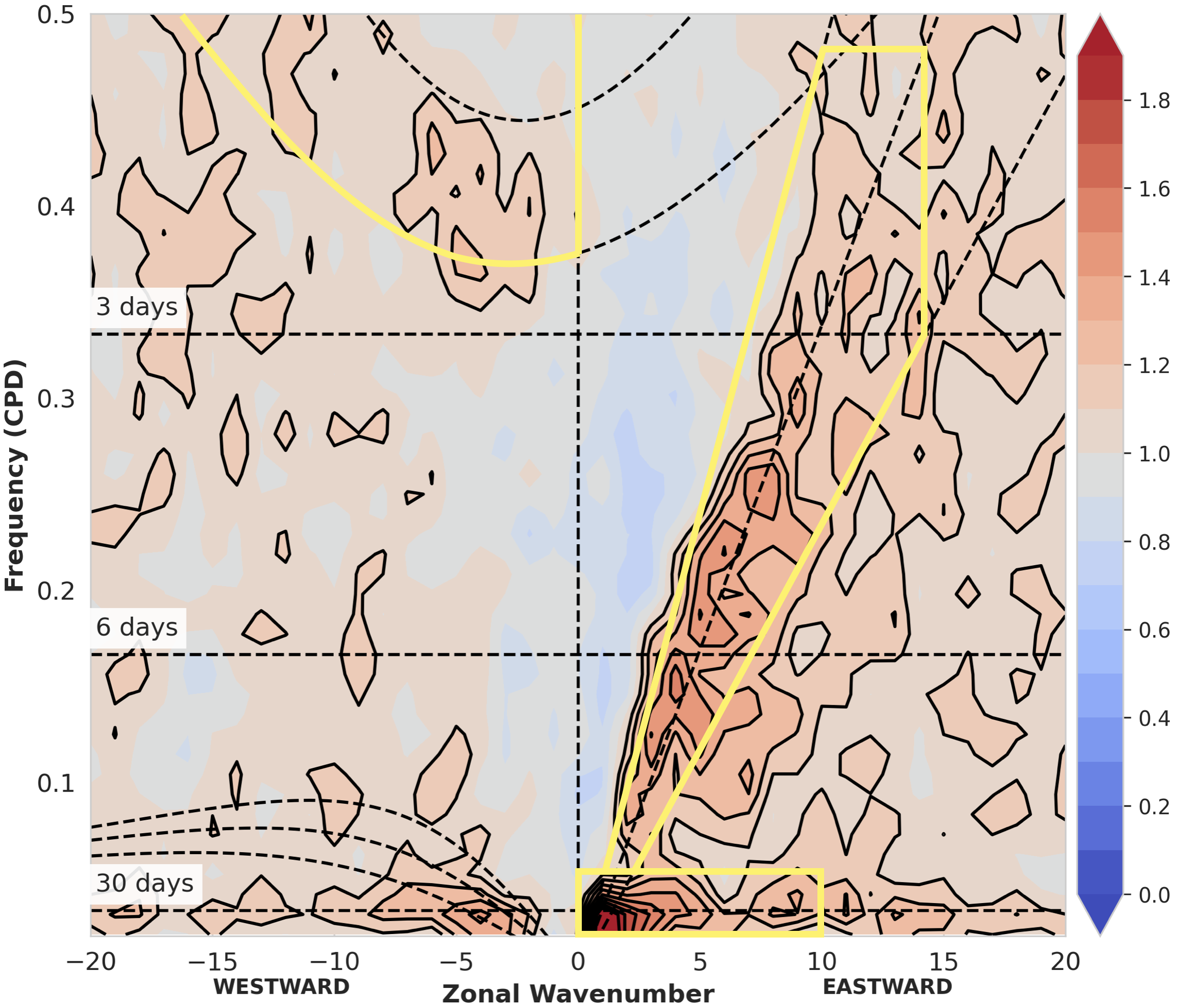}
        \caption{Observation}
        \label{fig:results-wk-obs}
    \end{subfigure}
    ~ 
    \begin{subfigure}{0.49\textwidth}
        \includegraphics[width=\textwidth]{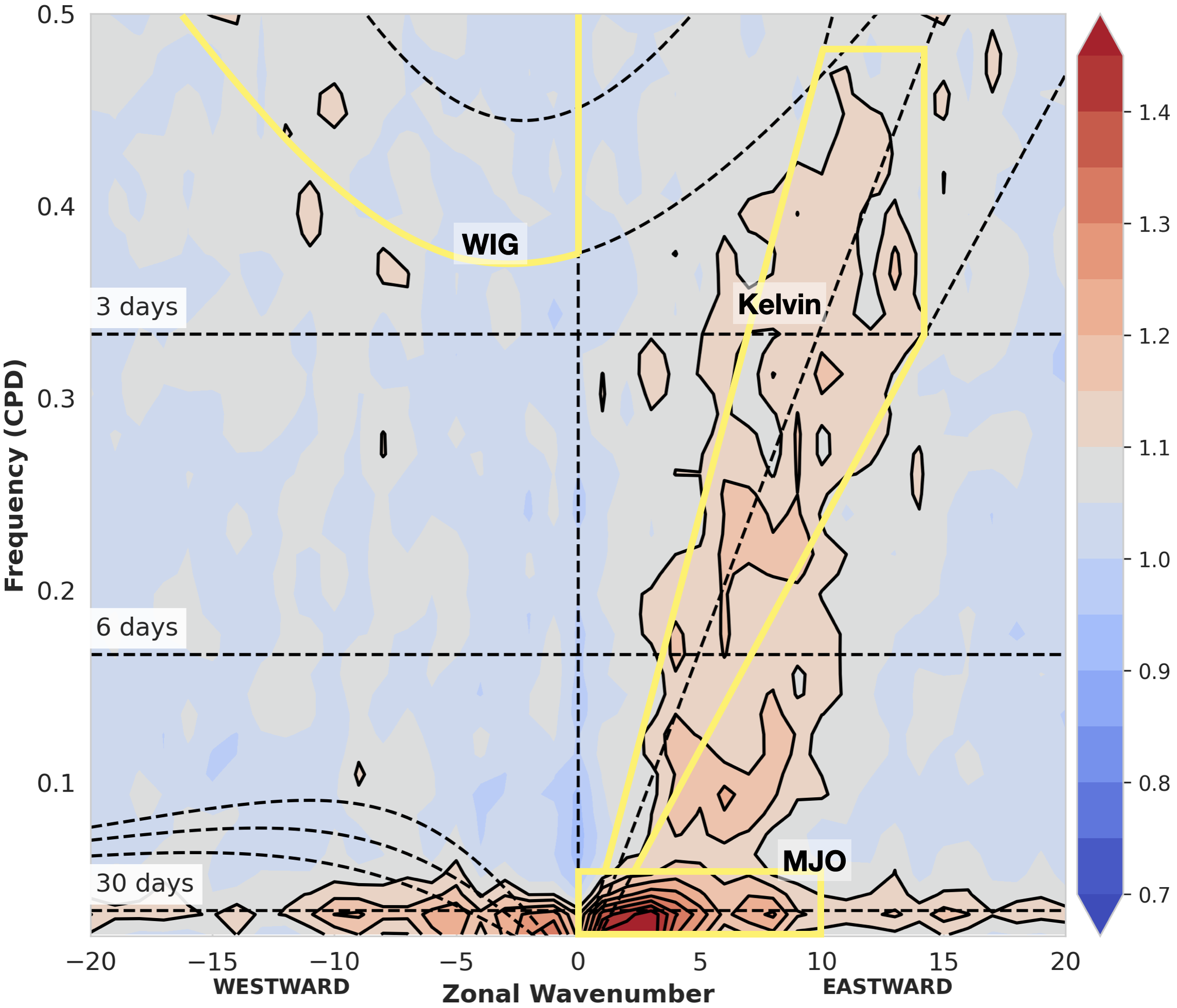}
        \caption{DiffObs}
        \label{fig:results-wk-diffobs}
    \end{subfigure}
    \caption{Symmetric / Background Wheeler--Kiladis space-time spectra between $15^\circ$N and $15^\circ$S. The individually highlighted \textcolor{black!20!yellow}{regions} correspond to where the Madden‐-Julian oscillation (MJO), Kelvin and westward inertio-gravity (WIG) waves are expected to be found.}
    \label{fig:results-wk}
\end{figure}

\cref{fig:results-hov,fig:results-wk} illustrate our key findings and capabilities. We first compare how convectively coupled equatorial waves (averaged between $5^\circ$S and $5^\circ$N) propagate through longitude and time relative to observations with a Hovm\"{o}ller Diagram \citep{hovmoller1949trough} --- a preferred domain diagnostic --- in \cref{fig:results-hov}. Qualitatively, a reassuring superposition of eastward- and westward-propagating tropical disturbances are generated at appropriate longitudes, modulated by a large-scale envelope of slow, eastward moving variability characteristic of the Madden‐-Julian oscillation \cite[MJO,][]{madden1994observations}, at its expected location spanning the Indian Ocean ($50$--$60^\circ$E) to West-Central Pacific ($180^\circ$E). Encouraging East Pacific Wavetrain \citep[EPW,][]{zhou2012east} variability is also found, alongside some artifacts, such as a dateline discontinuity at 180$^\circ$E/W and a bias towards too much time-mean precipitation generated between (additional comments in \cref{app:experiments}).



Our second analysis (\cref{fig:results-wk}) examines the dispersion relationships revealed in the wavenumber--frequency domain, following the methods in \cite{wheeler1999convectively,maier2018wk}. We generate \SI{80}{yrs} of data on one-day intervals, initially conditioning \SI{1}{yr} rollouts on Jan 1 for years 2017–2021 and sample with perturbed noise. Temporally concatenating the results within $15^\circ$N/S of the equator, we perform spectral analysis to construct a Wheeler--Kiladis diagram, utilizing \SI{96}{\day} windows with a \SI{65}{\day} overlap to isolate the significant spectral peaks. 

As a baseline, \cref{fig:results-wk-obs} illustrates the equatorially symmetric signal-to-noise for observations from the 5 years of test data. Key features, familiar to domain scientists, include the dominant power and east-to-west asymmetry on intraseasonal time scales (periodicity longer than 30 days), notably highlighting the MJO, as well as elevated power for an eastward-propagating convectively coupled Kelvin wave \citep{straub2002observations}, spanning wavenumbers $1$--$14$ with periods ranging from $2.5$ to $25$ days, and exhibiting a quasi-linear dispersion relationship \citep{kiladis2005zonal}. 

Encouragingly, DiffObs reproduces both of these dominant observed features (\cref{fig:results-wk-diffobs}): the spectral signal of generated power also occurs on intraseasonal time scales, i.e., timescales longer than $30$ day periodicity, and across a band of spatial (zonal) wavenumbers $0$--$9$ consistent with a planetary scale, eastward moving mode of variability. Meanwhile, on shorter time scales, the model also generates a moist Kelvin wave spectral power maximum  with a qualitatively correct dispersion relationship. Despite other imperfections, such as a tendency for the model to generate too much variability at all wavelengths (\cref{fig:app-wk-3,fig:app-wk-5}), and an under-representation of power within westward moving tropical wave classes, these are impressive preliminary results. Further discussion of the antisymmetric component and log power spectra can be found in \cref{app:wk}. Altogether, these results demonstrate the nontrivial ability to learn and autoregressively generate complex multi-scale tropical weather patterns over extended temporal scales, from precipitation data alone.

\section{Conclusion}
We have demonstrated an autoregressive, univariate, machine learning diffusion model for predicting daily-accumulated precipitation based only on the previous day's data. The model produces stable long rollouts that exhibit a realistic spectrum of tropical wave modes, including the Madden--Julian oscillation, whose variance dominates on intraseasonal time scales, and which is notoriously difficult to simulate realistically in physics-based models \citep{Zhangetal20}. Acknowledging secondary issues and biases, our overall findings hold promise for the next generation of global diffusion models, emphasizing the impact of training on increasingly sparse and differentiated observations of the world for subseasonal and climate prediction applications.

One caveat of this work is that the IMERG data we use is a level 3 product that integrates many separate data sources, including ERA5 \citep{hersbach2020era5}. Nonetheless, we feel this work takes a bold step away from reanalysis and future work could extend to even lower-level satellite products. Moreover, we would like to explore approaches to explicitly capture the temporal distributions, e.g., use multiple timesteps as the input condition to estimate $p\left(\mathbf{x}_t \vert \mathbf{x}_{t-1},\mathbf{x}_{t-2}\right)$, as well as more directly compare to numerical weather prediction and reanalysis data, e.g., the IFS and ERA5.

\subsubsection*{Acknowledgments}
This work was supported by NVIDIA through an internship with the Climate Simulation Research Group. This work was also supported by NSF Grant No. 2019758, \textit{AI Institute for Research on Trustworthy AI in Weather, Climate, and Coastal Oceanography (AI2ES).}

\bibliography{main}
\bibliographystyle{main}

\appendix
\section*{Appendix}

\section{Diffusion Details} \label{app:diffusion}
Diffusion methods are defined by separate forward and backward processes as represented by stochastic differential equations (SDEs). At a high level, these processes continuously increase or decrease the noise level of an input when moving forward or backward in time, respectively. Concretely, these SDEs evolve a sample, $\mathbf{x}$, to align with some data distribution, $p$, as it propagates through time \citep{karras2022elucidating,song2020score}. Leveraging a numerical solver we define a noise scheduler, $\sigma(t)$, to prescribe a given noise level at time $t$, typically as $\sigma(t) \propto \sqrt{t}$. 

\begin{figure*}[!ht]
    \centering
    \includegraphics[width=1\textwidth]{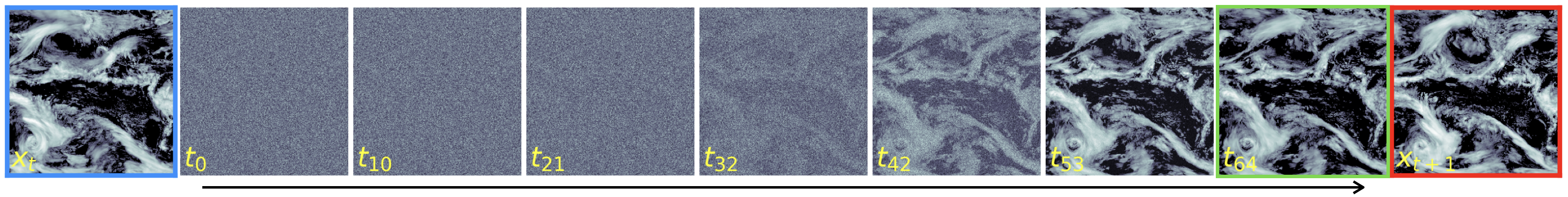} 
    \caption{Reverse diffusion of a cropped sample with the \textcolor{cyan}{input condition}, individual sampling steps ($t_0\rightarrow t_{64}$, inversely labeled), and the next time step \textcolor{black!10!green}{estimate} and \textcolor{red}{target} output.}
    \label{fig:diffusion-steps}
\end{figure*}

The \textit{forward} (drift-removed) SDE, as formulated by \cite{karras2022elucidating,mardani2024residual}, is expressed as
\begin{equation}\label{eq:diff-forward}
\mathrm{d} \mathbf{x} = \sqrt{2\dot{\sigma}(t) \sigma(t)}\, \mathrm{d}\omega_t,
\end{equation}
while the \textit{reverse}-time SDE \citep{anderson1982reverse}, sampled iteratively starting from $\mathbf{x}(T) \sim \mathcal{N} (\mathbf{0}, \sigma^2 \mathbf{I})$ for a large $T\dots0$ (illustrated in \cref{fig:diffusion-steps}), is defined as
\begin{equation}\label{eq:diff-backward}
\mathrm{d} \mathbf{x} = - 2\dot{\sigma}(t) \sigma(t) \nabla_\mathbf{x} \log p\left(\mathbf{x}; \sigma(t)\right) \, \mathrm{d}t + \sqrt{2\dot{\sigma}(t) \sigma(t)}\, \mathrm{d}\bar{\omega}_t.
\end{equation}
Here, $\dot{\sigma}(t)$ is the time derivative of $\sigma(t)$ and $\nabla_\mathbf{x} \log p(\mathbf{x}; \sigma)$ is the score function \citep{hyvarinen2005estimation}. The two terms in \cref{eq:diff-backward} are the deterministic component representing the probability flow ordinary differential equation (ODE) with noise degradation, and noise injection via the standard Wiener process (denoted by $\omega_t$), respectively.

The significance of the score function in \cref{eq:diff-backward} lies in its relevance to sampling from diffusion models. Notably, it has the intriguing characteristic of not relying on the typically intractable normalization constant of the underlying base distribution $p(\mathbf{x}; \sigma)$. Exploiting this independence, we can use a denoising method defined by a neural network that minimizes the expected $L_2$ loss as a substitute of the score function, i.e., $\nabla_\mathbf{x} \log p(\mathbf{x}; \sigma) = \left(D_\theta\left(\mathbf{x}, \sigma\right) - \mathbf{x}\right) / \sigma^2$.

Let $D_\theta$ be a denoising model that operates on a noisy input sample given $\mathbf{x}_t \in \mathbb{R}^{c \times h \times w}$ at sample time $t$ and noise level $\sigma$, where the previous state or condition, $\mathbf{x}_{t-1}$, is concatenated channel-wise to the input. Concretely, $\mathbf{x}_t$ is the target next-day sample and $\mathbf{x}_{t-1}$ is the previous day input condition. We therefore optimize $D_\theta$ in training using
\begin{equation} \label{eq:loss}
\min_{\theta} \mathbb{E}_{\mathbf{x}_{t,t-1}\sim p_{\mathrm{data}}} \mathbb{E}_{\sigma\sim p_\sigma} \mathbb{E}_{\mathbf{n}\sim\mathcal{N} \left(\mathbf{0},\,\sigma^2\mathbf{I}\right)} \left[\lambda(\sigma)\lVert D_\theta\left(\mathbf{x}_t + \mathbf{n}, \mathbf{x}_{t-1};\sigma\right)\,-\,\mathbf{x}_t\rVert^2_2\right],
\end{equation}
where the loss weight $\lambda(\sigma) = (\sigma^2 + \sigma^2_{\mathrm{data}}) / (\sigma \cdot \sigma_{\mathrm{data}})^2$, the noise level $\sigma$ follows a log-normal distribution $\ln(\sigma) \sim \mathcal{N}(-1.2,1.2^2)$, and $\sigma_\mathrm{data} = 0.5$.

The denoising model, $D_\theta$, with an underlying trainable network, $F_\theta$, is preconditioned following
\begin{equation}
D_\theta\left(\hat{\mathbf{x}}_{t},\mathbf{x}_{t-1}; \sigma\right) = c_{\mathrm{skip}}(\sigma)\hat{\mathbf{x}}_{t} + c_{\mathrm{out}}(\sigma) F_\theta \left(\left[c_{\mathrm{in}(\sigma)}\hat{\mathbf{x}}_{t}, \mathbf{x}_{t-1}\right]; c_{\mathrm{noise}}(\sigma)\right),
\end{equation}
where the noisy input $\hat{\mathbf{x}}_{t} = \mathbf{x}_{t} + \mathbf{n}$, and $c_{\mathrm{*}}(\sigma)$ \citep[Table 1]{karras2022elucidating} are preconditioning variables to scale and modulate the individual components. The previous timestep (condition) is concatenated channel-wise by $[\cdot]$ and $c_{\mathrm{noise}}(\sigma)$ is an additional latent condition for $F_\theta$.

To generate samples from our diffusion model, we leverage \cref{eq:diff-backward} with our trained denoising network $D_\theta$. Specifically, we iteratively solve this using the stochastic EDM sampler that combines a second-order deterministic ODE integrator with stochastic Langevin-like churn. The sampled output is used autoregressively as the condition for the next sample time step.

\section{Wheeler--Kiladis Space-Time Spectra} \label{app:wk}
In their seminal work, \cite{wheeler1999convectively} showed how fast and slowly oscillating atmospheric waves, some of which arise in idealized theories \citep{yanai1966stratospheric}, can be observed through spectral analysis of satellite-observed outgoing longwave radiation, including the Madden--Julian oscillation. This form of two dimensional spectral analysis has became one of the fundamental techniques for evaluating numerical models' representation of tropical waves that regulate predictability on synoptic to subseasonal timescales. Computing the Wheeler--Kiladis diagram reveals that DiffObs captures many, if not all, of the predominant observed modes and tropical wave signals. This could be viewed as a surprising property of a machine learning model trained only on a single variable, given that dynamical theories of such waves encompass several atmospheric state variables operating in concert --- e.g., the vorticity and divergence of the horizontal winds are required for modeling in the equatorial Rossby and Kelvin waves, respectively, while the MJO requires complex non linear advection and moisture effects. 

\begin{figure}
    \centering
    \begin{subfigure}{0.49\textwidth}
        \includegraphics[width=\textwidth]{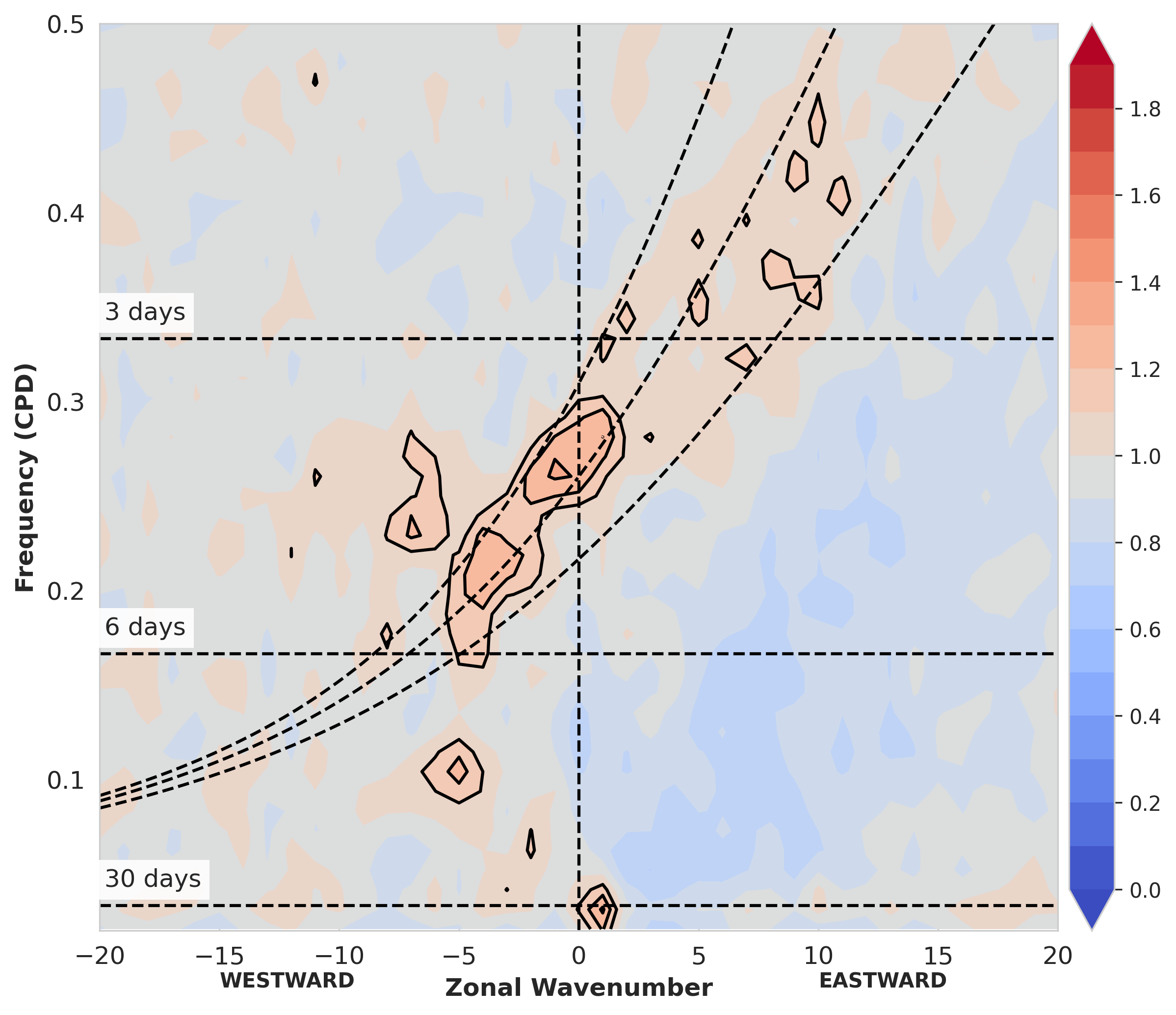}
        \caption{Antisymmetric / Background (\textbf{Obs})}
        \label{fig:app-wk-0}
    \end{subfigure}
    ~ 
    \begin{subfigure}{0.49\textwidth}
        \includegraphics[width=\textwidth]{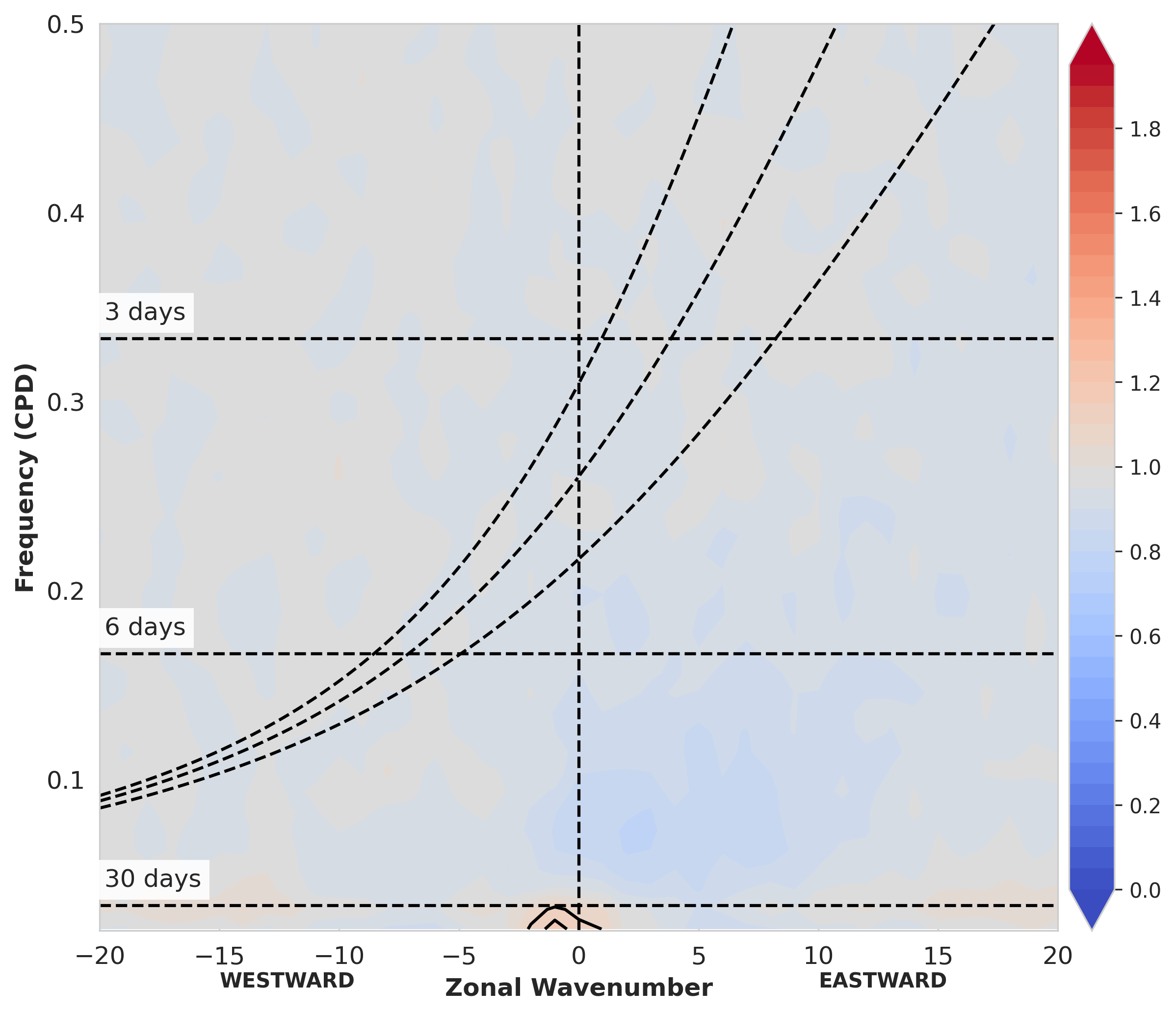}
        \caption{Antisymmetric / Background (\textbf{DiffObs})}
        \label{fig:app-wk-1}
    \end{subfigure}

    \vspace{0.05\linewidth} 
    
    \begin{subfigure}{0.49\textwidth}
        \includegraphics[width=\textwidth]{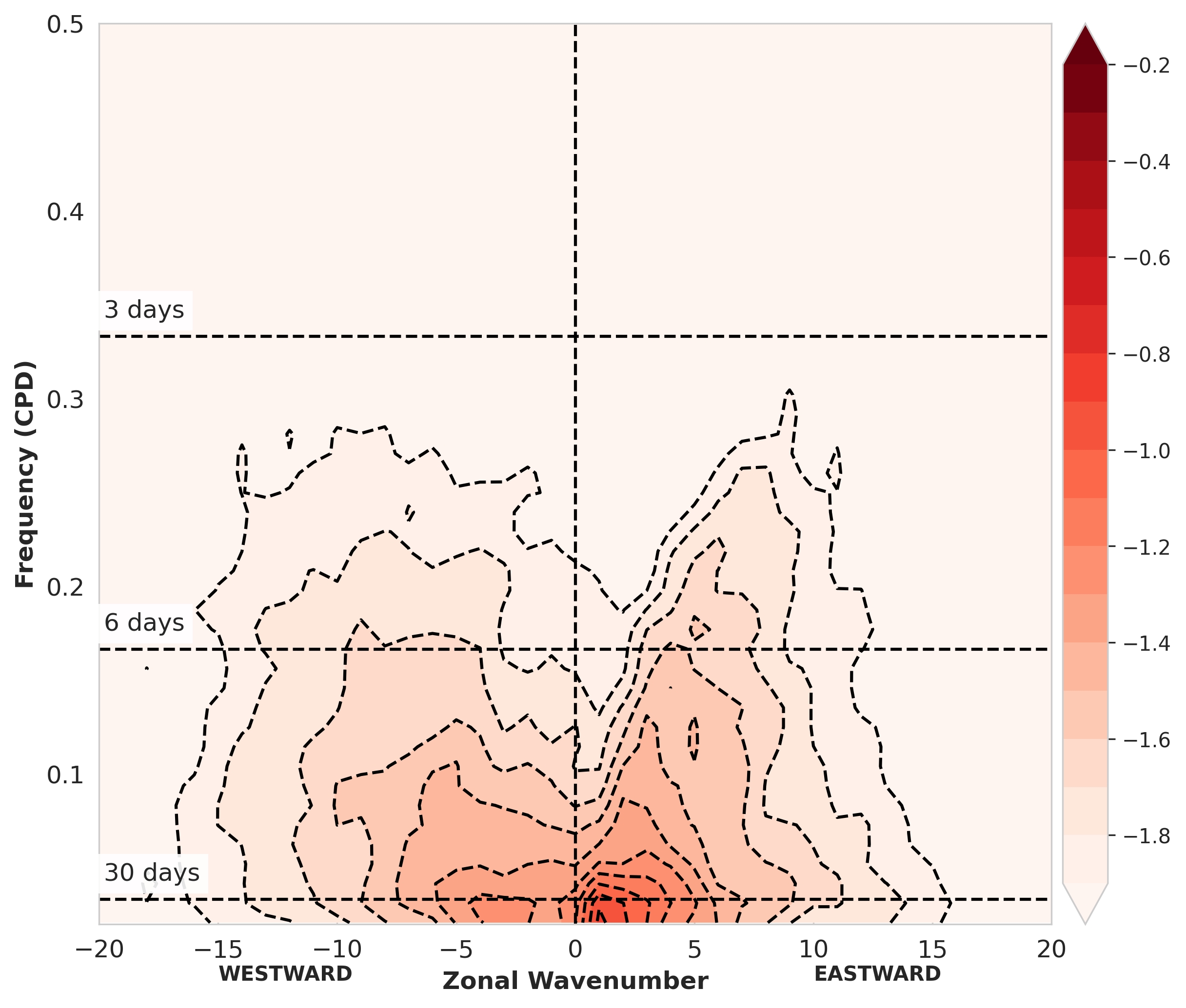}
        \caption{Symmetric Raw Log Power (\textbf{Obs})}
        \label{fig:app-wk-2}
    \end{subfigure}
    ~ 
    \begin{subfigure}{0.49\textwidth}
        \includegraphics[width=\textwidth]{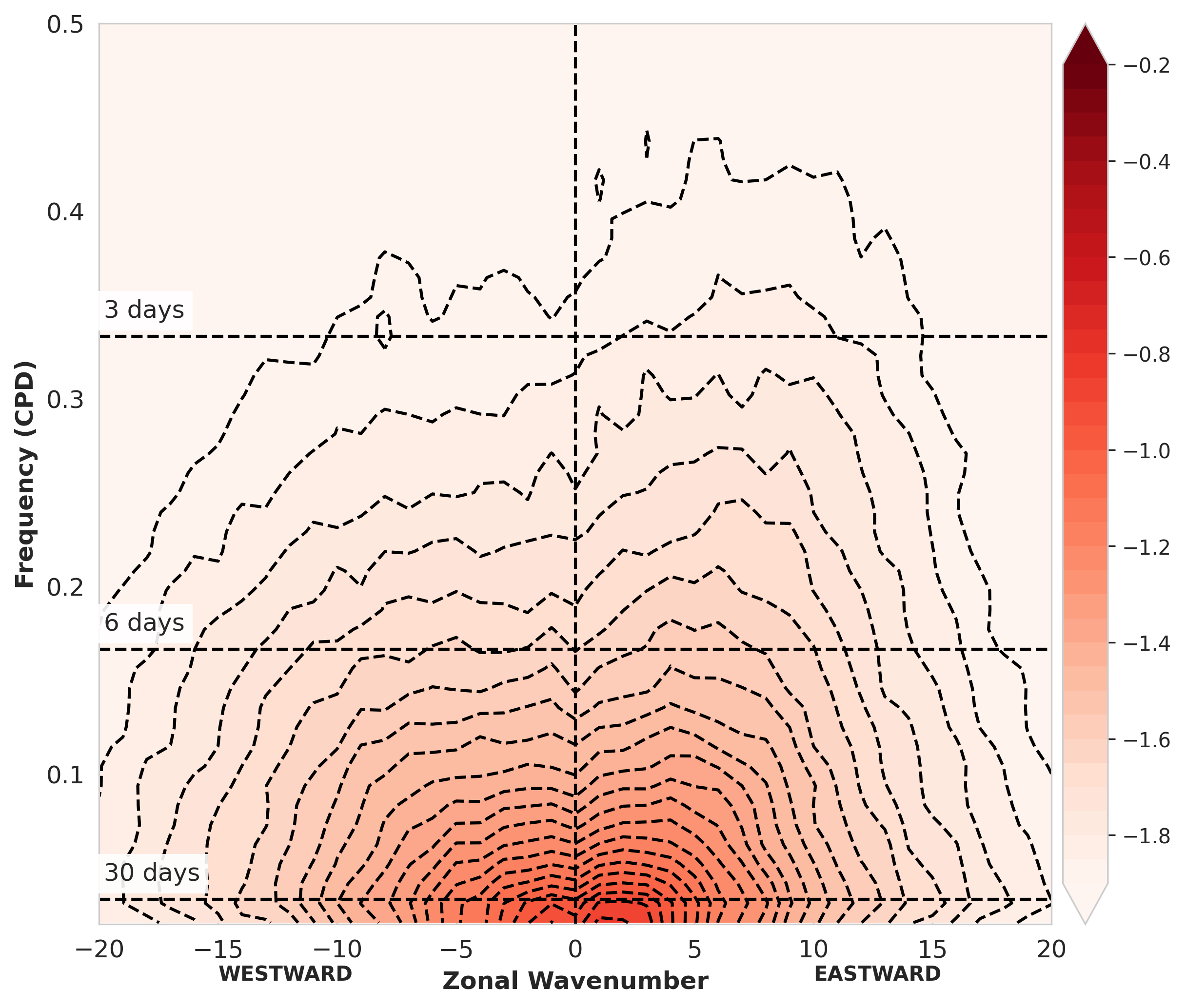}
        \caption{Symmetric Raw Log Power (\textbf{DiffObs})}
        \label{fig:app-wk-3}
    \end{subfigure}

    \begin{subfigure}{0.49\textwidth}
        \includegraphics[width=\textwidth]{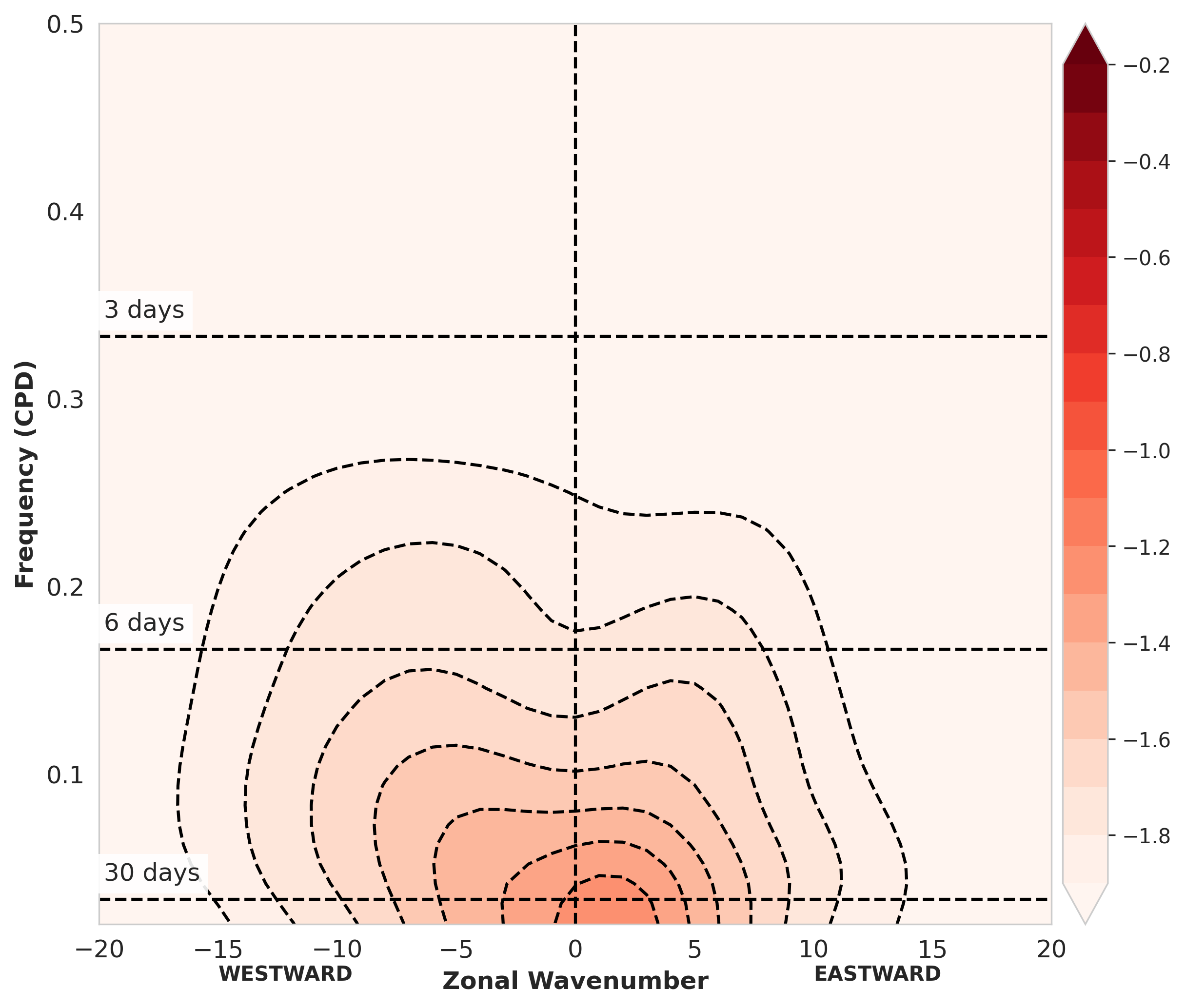}
        \caption{Background Log Power (\textbf{Obs})}
        \label{fig:app-wk-4}
    \end{subfigure}
    ~ 
    \begin{subfigure}{0.49\textwidth}
        \includegraphics[width=\textwidth]{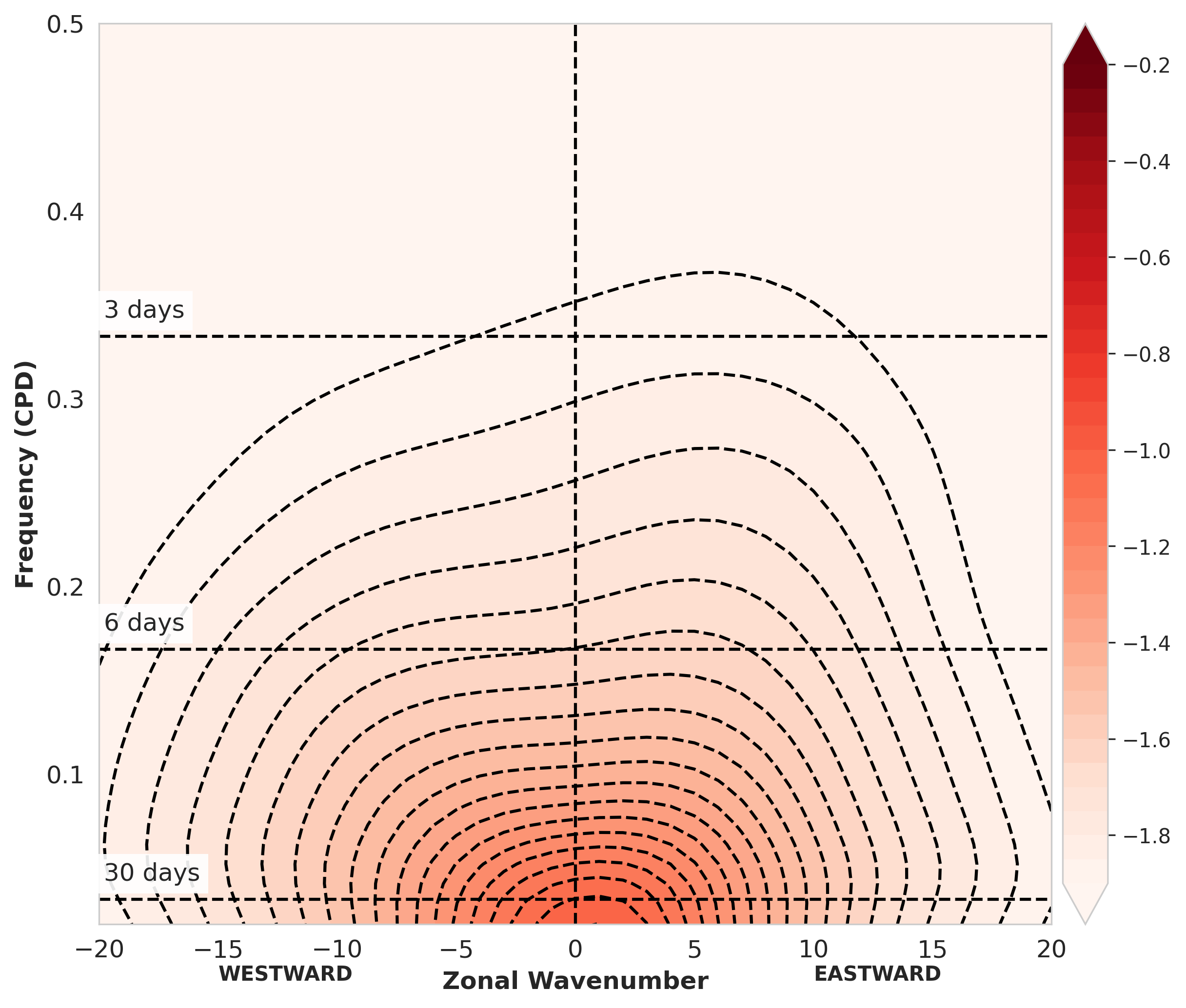}
        \caption{Background Log Power (\textbf{DiffObs})}
        \label{fig:app-wk-5}
    \end{subfigure}

    \caption{Additional Wheeler--Kiladis components and power spectra of observations (left column) and model output (right column) that support \cref{fig:results-wk}.}
    \label{fig:app-wk}
\end{figure}

Our main finding is the discovery of Kelvin wave and MJO spectral signals within the signal-to-noise ratio of the equatorially-symmetric component of the space-time spectra (see \cref{fig:results-wk-diffobs}). However, this is only one view of the analysis, and we glean more details from the additional components of the analysis. Specifically, in the antisymmetric component of observations (\cref{fig:app-wk-0}), there is evidence of a mixed Rossby-gravity (MRG) and eastward inertio-gravity (EIG) waves that are not apparent in \cref{fig:app-wk-1}. In \cref{fig:app-wk-3,fig:app-wk-5}, we show the intermediate computations, where the background spectra of model output is more similar to the raw power spectra than that of the observations. This suggests that DiffObs has a strong background signal that is similar to red-noise and that overall DiffObs generates too much variance.

\section{Short-Term Predictability}
While short-term predictability is not the focus of this work, we find it is important to assess for comprehension and to be relatable to prior work. Therefore, we evaluate quantitative model predictability using the Root-Mean-Squared Error \citep[RMSE,][]{price2023gencast} relative to forecasts derived by the observations and the Fractions Skill Score \citep[FSS,][]{roberts2008scale,roberts2008assessing}. While the Continuous Ranked Probability Score (CRPS) and Brier Skill Score (BSS) are common metrics to evaluate forecast performance, we currently do not have other datasets to make for a meaningful comparison. As such, FSS is defined for a given neighborhood size as
\begin{equation}
\mathrm{FSS} = 1 - \frac{\frac{1}{n}\sum_n\left(P_y - P_t\right)^2}{\frac{1}{n}\left[\sum_n P^2_y + \sum_n P^2_t\right]},
\end{equation}
where $P_y$ is the forecast fraction and $P_t$ is the target observed fraction (exceeding a certain threshold), and $n$ is the number of spatial windows over the domain. We define the ensemble mean RMSE similar to \cite{price2023gencast} as
\begin{equation}
\mathrm{RMSE} = \sqrt{\frac{1}{M}\sum_m\frac{1}{\lvert G \rvert}\sum_i a_i \left(t_{i,m} - \bar{y}_{i,m}\right)^2},
\end{equation}
where $y_{i,m}^n$ denotes the $n \in N$ ensemble for the $m \in M$ forecast for a lead time at a latitude and longitude indexed by $i \in G$, $t_{i,m}$ is the target observation, and $\bar{y}_{i,m} = 1 / N \sum_n y^n_{i,m}$ is the ensemble mean of model predictions. We use a latitude weighting derived from the area mean on a sphere, normalized to have unit mean as defined by
\begin{equation}
a_i = \frac{\cos\left(\mathrm{lat}\left(i\right)\right)}{\frac{1}{N_{\mathrm{lat}}} \sum_j \cos\left(\mathrm{lat}\left(j\right)\right)}.
\end{equation}

We compare individual scores to the \textit{persistence forecast} taken as the initial condition repeated over the forecast duration (14-days) as well as \textit{climatology} found by a two week average window of mean daily conditions for years 2000--2022. Additionally, we compute the ensemble mean errors individually at the midlatitudes (between $30^\circ$N/S and $60^\circ$N/S) and tropics (between $23.5^\circ$N and $23.5^\circ$S) without any latitude weighting. We also show the deterministic error and the ensemble spread as the square root of $(n+1)/n$ times the average over all forecasts of the ensemble variance. 

\begin{figure}[t!]
    \centering
    \begin{subfigure}{0.49\textwidth}
        \includegraphics[width=\textwidth]{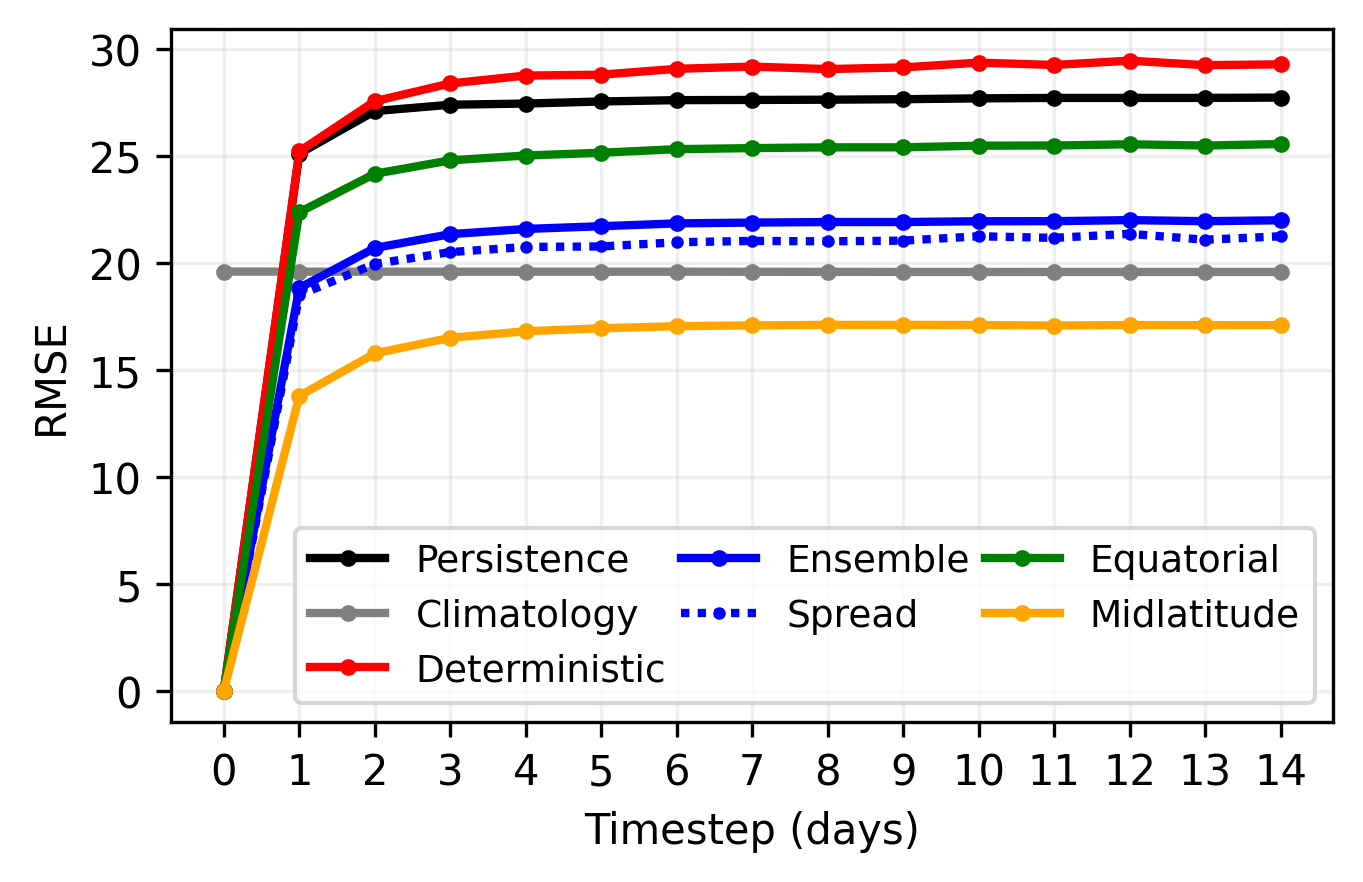}
        \caption{RMSE}
        \label{fig:metrics-rmse}
    \end{subfigure}
    ~ 
    \begin{subfigure}{0.49\textwidth}
        \includegraphics[width=\textwidth]{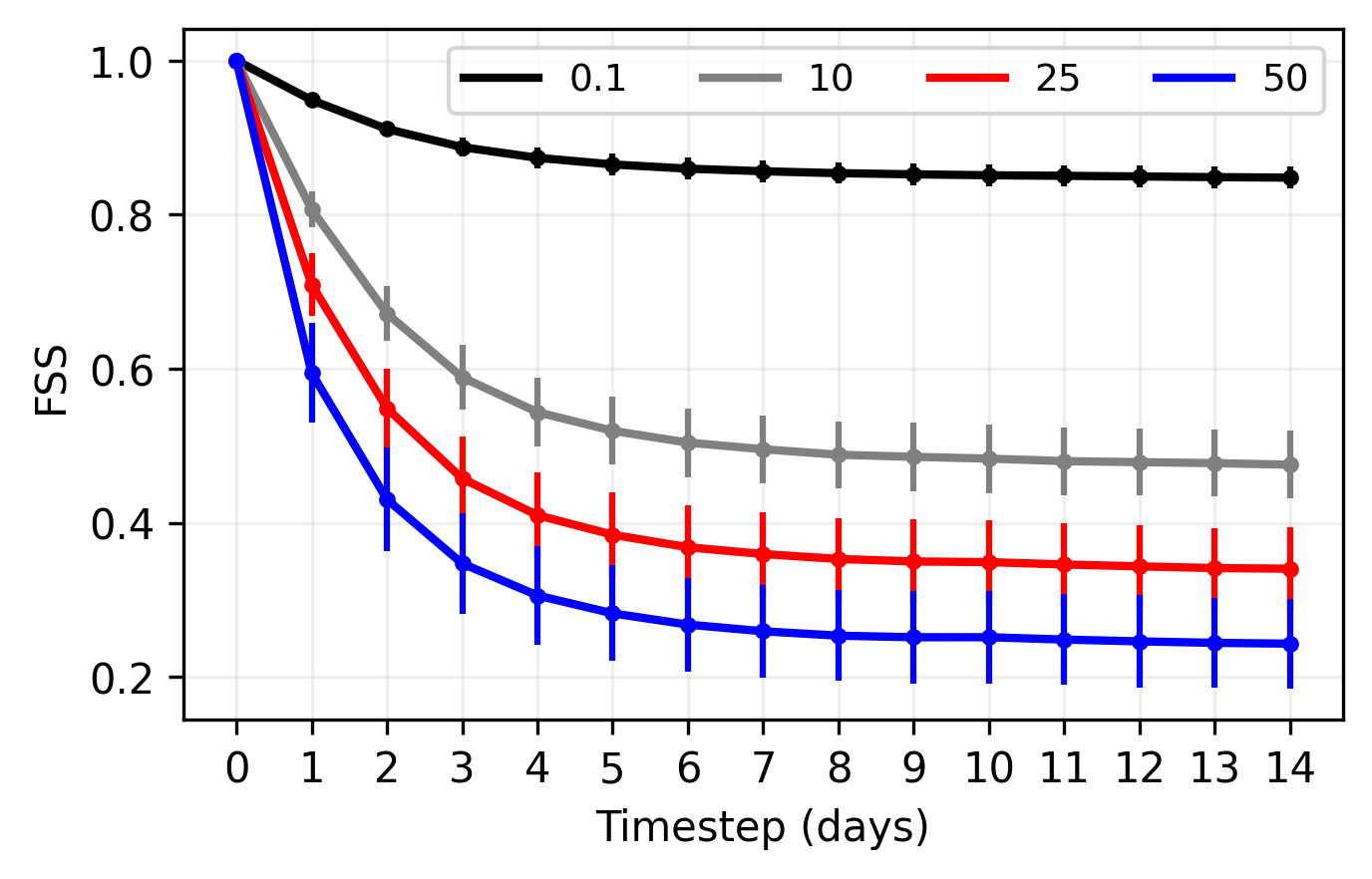}
        \caption{FSS}
        \label{fig:metrics-fss}
    \end{subfigure}
    \caption{Short-term predictability for 14-day forecasts, initiated from each day between Jan 1, 2020 and Sept 15, 2021 with 5 ensemble members. Panel (a) shows the RMSE and (b) shows the ensemble mean FSS with different thresholds (\si{\milli\meter/\day}) at a $8.4^\circ$ neighborhood (21 pixels).}
    \label{fig:metrics}
\end{figure}

\cref{fig:metrics-rmse} shows the forecast errors plateau quickly after a 3-day lead time, yet there is consistency with the persistence and deterministic errors approaching the ratio of $\sqrt{2}$ with lead time for climatology and the ensemble, respectively. Even though the ensemble error is better than persistence, given this consistency and that the model is under-dispersive (spread $<$ error), we can deduce that the ensemble mean is relatively poor. By computing errors at varying latitudes, we see the greatest error with the estimations in the tropics, where there is also a bias toward high precipitation values.

In evaluating our model using FSS, various neighborhood sizes were considered (not shown within), revealing that the $8.4^\circ$ neighborhood size effectively captures large-scale atmospheric structure. The outcomes of experimenting with different thresholds are presented in \cref{fig:metrics-fss}. Notably, lower thresholds have higher skill, which continuously decreases with high-valued estimates, and the highest skill is at early lead times (out to 5 days), where skill plateau thereafter as seen with RMSE.

\section{Additional Experiments} \label{app:experiments}

In our experiments, we observe a dateline discontinuity at $180^\circ$E/W, evident down the center of the Hovm\"{o}ller in \cref{fig:results-hov} (right). Ideally, we would like our model to be consistent around the globe and allow for periodic wave propagation. We aim to address this by modifying our network architecture and by including additional conditions, as outlined in \cref{app:add-mods}. However, we find the changes to be suboptimal, showing in \cref{app:add-results} further inconsistencies and worse performance relative to the baseline model.

\subsection{Modifications} \label{app:add-mods}

For our network to have rotational equivariance, we modify the convolutions to use circular-padding in the zonal direction and zero-padding in meridional direction (due to the cylindrical structure of our data). This effectively removes any spatial bindings given by the dateline. As such, we include a two-channel static condition (concatenated channel-wise to the existing condition) of $\cos(\mathrm{lon})$ and $\sin(\mathrm{lon})$, repeated over the meridional directions. These additional conditions, spatially aligned with the input, \textit{should} maintain spatial coherence. 

In addition to the padding and coordinate conditions, we also include the zonal average of the cosine of solar zenith angle as a function of the condition date and latitudes to account for temporal variability. We compute this for each latitude, $\phi$, at time $t$ as,
\begin{equation}
    \cos \theta_s = \sin \phi \sin \delta + \cos \phi \cos \delta \cos h,
\end{equation}
where $\delta$ is the declination of the sun and $h$ is the hour angle. Given that our data is daily-accumulated, we integrate time by zonally averaging at UTC+0 and repeat the value for each latitude. The result is again concatenated channel-wise and \textit{should} be effective to provide seasonal cycle conditioning. It is important to note that during training, we use the date associated with the condition (i.e., the previous timestep), iterating $\cos \theta_s$ in time during a rollout.

\subsection{Preliminary Results} \label{app:add-results}

\begin{figure}[t!]
    \centering
    \begin{subfigure}{0.405\textwidth}
        \includegraphics[width=\textwidth]{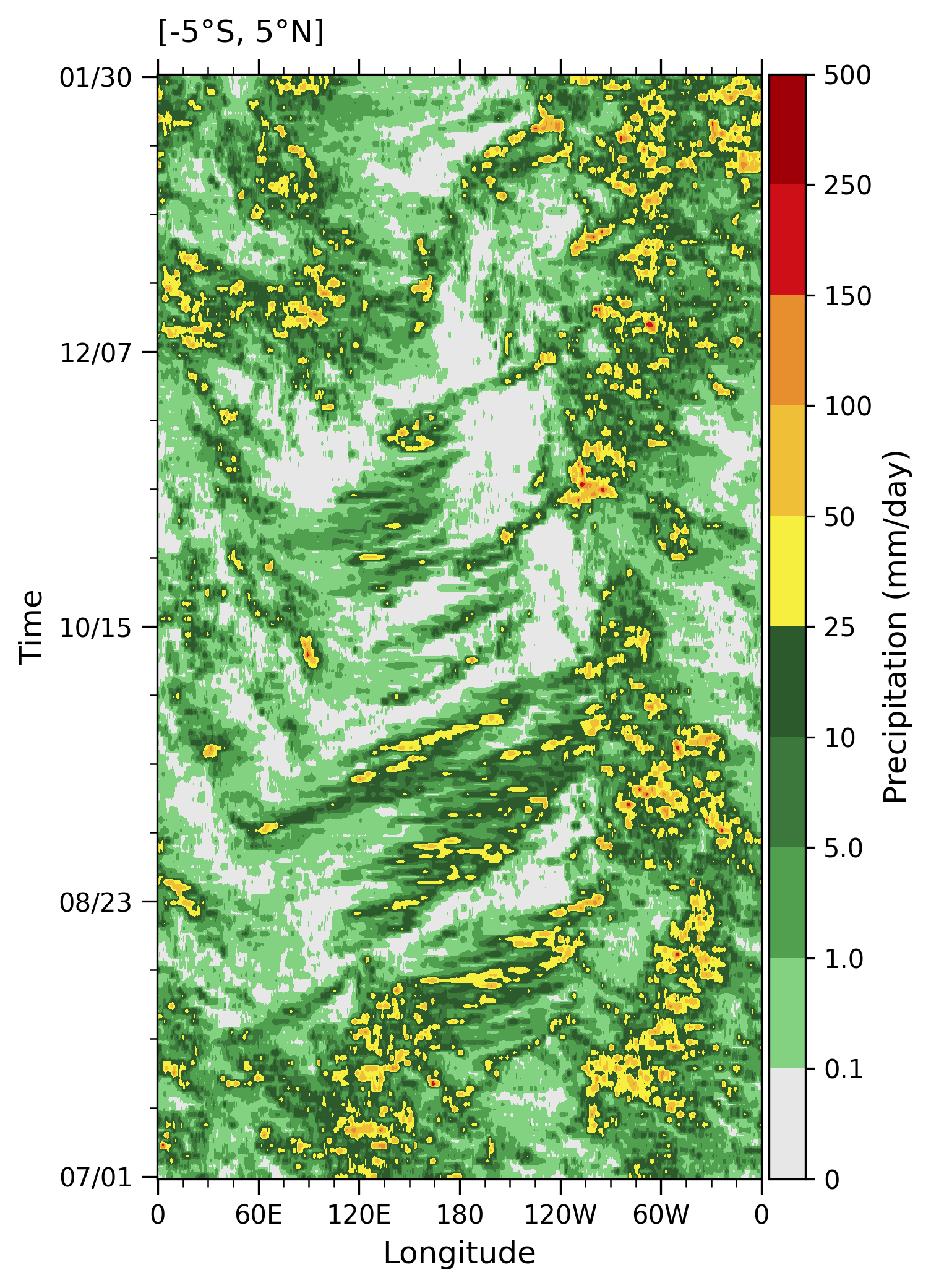}
        \caption{Hovm\"{o}ller}
        \label{fig:add-hov}
    \end{subfigure}
    ~ 
    \begin{subfigure}{0.55\textwidth}
        \includegraphics[width=\textwidth]{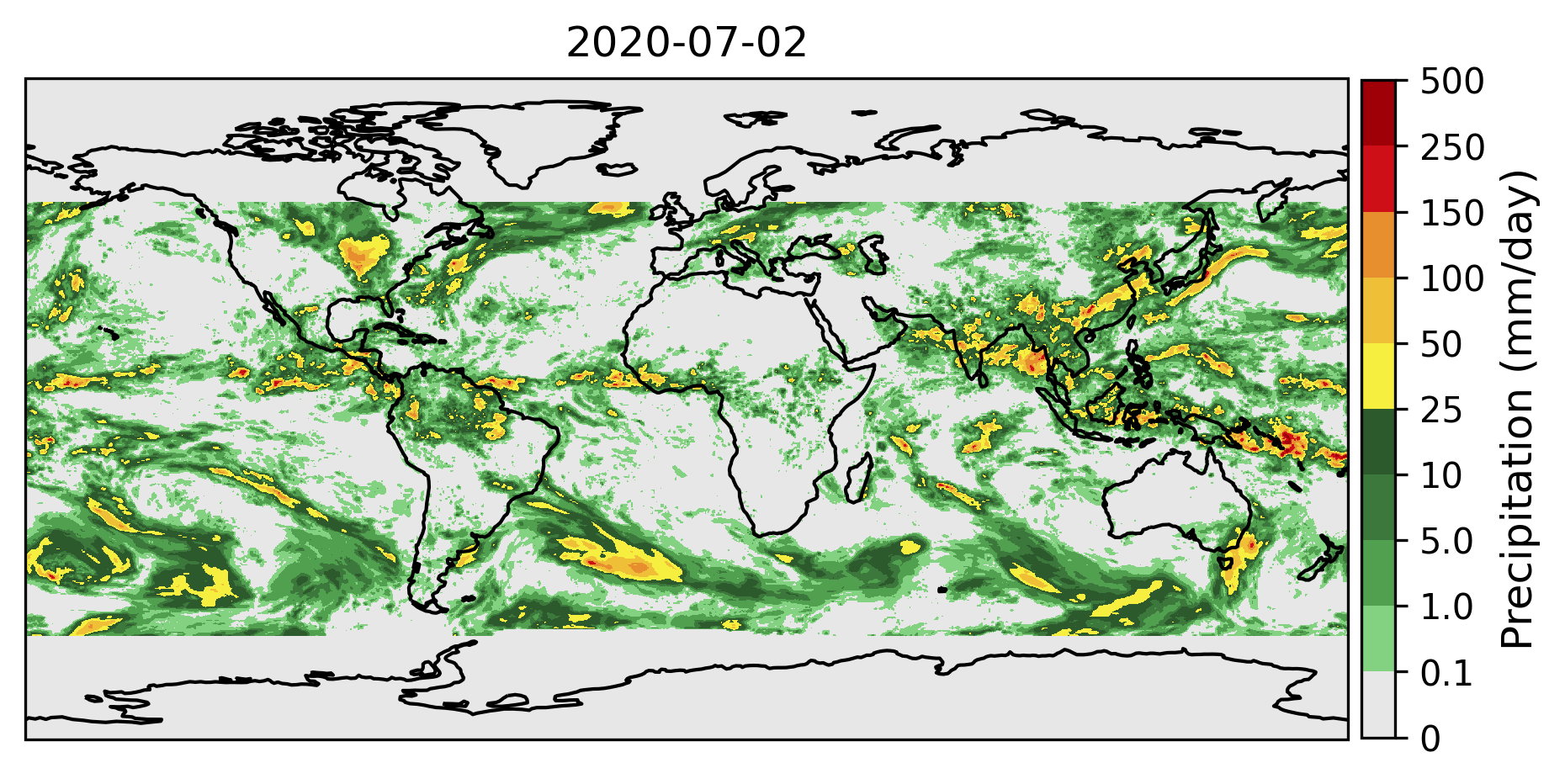}
        \includegraphics[width=\textwidth]{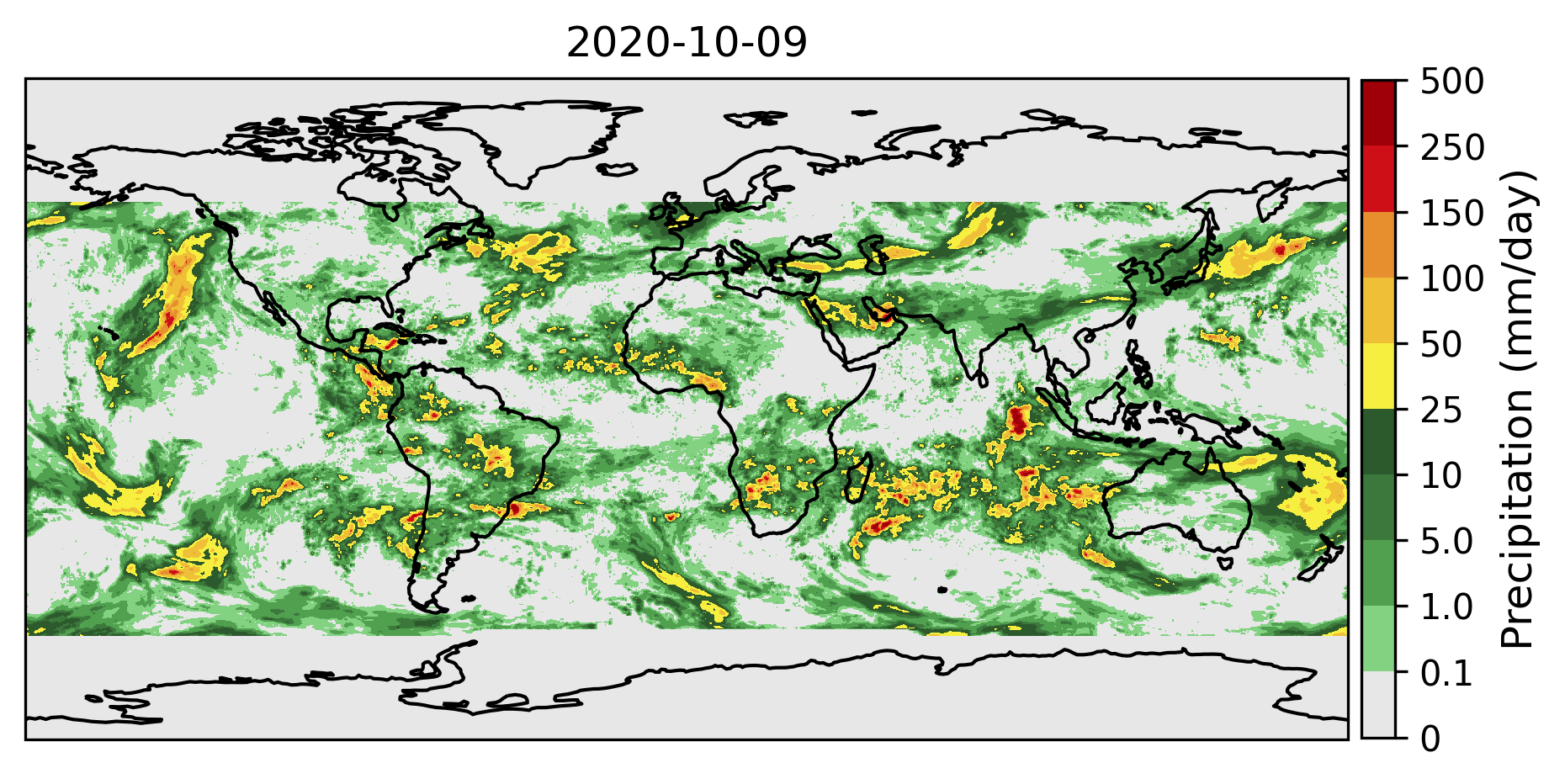}
        \caption{Individual Output Samples}
        \label{fig:add-samples}
    \end{subfigure}
    \caption{Experimental DiffObs output (\cref{app:experiments}), generated autoregressively when initially conditioned on July 1, 2020. Panel (a) is the Hovm\"{o}ller between $5^\circ$S to $5^\circ$N and (b) are individual output samples with their corresponding steps shown by the date.}
    \label{fig:add}
\end{figure}

We train our updated model with the same hyperparameters and training specifics detailed in \cref{sec:method} and repeat similar evaluations from \cref{sec:experiments}. While the periodicity is preserved across the dateline, we find the results to inadequately represent the atmospheric dynamics. The most salient representation of this is illustrated in \cref{fig:add-hov} when comparing to observation (\cref{fig:results-hov}, left). Notably, no landmasses are identified, eastward-propagating waves traverse the dateline, and oscillating wave signals are not captured. While it is not abundantly clear as to why these modifications yield worse results, we note that there should be careful considerations when iterating on future work.

\end{document}